\documentclass[aps,prl,reprint,groupedaddress, floatfix,nolongbibliography]{revtex4-2}
\usepackage{aas_macros}
\usepackage{orcidlink}
\usepackage{graphicx}
\usepackage{amsmath}

\begin{document}

\title{\textbf{Gravothermal Collapse: Robust Against Baryonic Feedback}}%

\author{Demao Kong\orcidlink{0000-0003-1723-8691}}
\email{dkong012@ucr.edu}
\affiliation{Center for Experimental Cosmology \& Instrumentation, Department of Physics \& Astronomy, University of California, Riverside, CA 92521, USA}

\author{Hai-Bo Yu\orcidlink{0000-0002-8421-8597}}
\email{haiboyu@ucr.edu}
\affiliation{Center for Experimental Cosmology \& Instrumentation, Department of Physics \& Astronomy, University of California, Riverside, CA 92521, USA}

\date{\today}

\begin{abstract}
We perform a stress test of gravothermal collapse in self-interacting dark matter (SIDM) halos under baryonic feedback using a semi-analytical oscillating-potential model in controlled N-body simulations. For high-concentration halos, where the SIDM thermalization timescale is short, gravothermal collapse is only mildly delayed and never stalled, even under extremely strong feedback. In contrast, the collapse of a median-concentration halo can be significantly delayed, but it resumes once feedback ceases. The final density profile of such halos depends sensitively on the episodic feedback history, producing a broad diversity in central densities. Our findings further establish gravothermal collapse as a defining prediction of dark matter self-interactions.

\end{abstract}

\maketitle

\textit{\textbf{Introduction--}}In the standard model of cosmic structure formation, dark matter consists of cold, collisionless particles. Within this cold dark matter (CDM) framework, dark matter halos generically develop steep central density cusps, in tension with the shallow density cores inferred in many dwarf galaxies~\cite{Flores:1994gz,Moore:1994yx,deBlok:2001rgg,Gentile:2004tb,Oh:2015xoa,Bullock:2017xww}. Baryonic feedback associated with galaxy formation can inject energy into the halo and potentially transform cuspy profiles into shallow cores~\cite{Navarro:1996bv,Governato:2009bg,Pontzen:2011ty,DiCintio:2013qxa,Read:2015sta,Chan:2015tna,2023MNRAS.520..461J,Cruz:2025wfz}, alleviating the traditional core--cusp tension within the CDM framework. However, more recent studies have revealed a broad diversity of dark matter distributions in galaxies~\cite{KuziodeNaray:2009oon,Oman:2015xda,Read:2016xbf,Kamada:2016euw,Creasey:2017qxc,Ren:2018jpt}, ranging from shallow cores to dense cusps among halos of similar mass, which remains challenging to reproduce in CDM~\cite{Kaplinghat:2019dhn,Santos-Santos:2019vrw,Zentner:2022xux,Jia:2026ocr,Sales:2022ich}.

Moreover, several dense and compact perturbers have been identified through observations of strong gravitational lensing systems \cite{Vegetti:2010,Vegetti:2012mc,Sengul:2021lxe,Ballard:2023fgi,Minor:2025,Tajalli:2025qjx,Powell:2025rmj}, stellar streams~\cite{Bonaca:2018fek,Nibauer:2025ezn}, and satellite galaxies \cite{Penarrubia:2024vms}. If interpreted as dark matter (sub)halos, these systems possess density concentrations substantially higher than those typically expected in CDM \cite{Minor:2020hic,Enzi:2024ygw,Despali:2024ihn,Yu:2025tmp,Vegetti:2026mmx}. For such objects, baryonic feedback that creates cores would further exacerbate the discrepancy between observations and CDM predictions. 

In the self-interacting dark matter (SIDM) framework, dark matter distributions can be significantly more diverse than those predicted in CDM; see Refs.~\cite{Tulin:2017ara,Adhikari:2022sbh} for reviews. Dark matter self-interactions drive gravothermal evolution, causing halos to evolve from a core-expansion phase~\cite{Spergel:1999mh} to a core-collapse phase~\cite{Balberg:2002ue}, characterized by shallow and steep density profiles, respectively. SIDM can naturally produce both low- and high-density halos by ``amplifying'' the intrinsic scatter in halo concentration~\cite{Essig:2018pzq,Sameie:2019zfo,Zeng:2021ldo,Nadler:2023nrd}, as demonstrated in cosmological zoom-in simulations~\cite{Turner:2020vlf,Correa:2022dey,Yang:2022mxl,Fischer:2023lvl,Nadler:2023nrd,Nadler:2025jwh,Despali:2025koj,Engelhardt:2026dpj}. 

Since gravothermal collapse is a defining feature of SIDM, confirming or refuting its observational signatures provides an important avenue for understanding the nature of dark matter. For example, core-collapsed SIDM (sub)halos may explain the high densities inferred for the perturbers discussed previously \cite{Nadler:2023nrd,Zhang:2024fib,Kong:2025sqx,Li:2025kpb,Yu:2025tmp}, dense spiral galaxies \cite{Roberts:2024uyw,Kong:2025irr} and Milky Way satellite galaxies~\cite{Zavala:2019sjk,Nishikawa:2019lsc,Sameie:2019zfo,Correa:2020qam,Slone:2021nqd,Fischer:2026ryr}. SIDM gravothermal collapse may also provide a mechanism for seeding supermassive black holes in the early Universe~\cite{Balberg:2001qg,Pollack:2014rja,Feng:2020kxv,Feng:2021rst,Gad-Nasr:2023gvf,Jiang:2025jtr,Feng:2025rzf,Shen:2025evo,Roberts:2025poo,Gu:2026zzq}. Most of these studies are based on SIDM-only simulations or semi-analytical modeling with static baryonic potentials. However, baryonic potentials shaped by feedback processes are often neither static nor adiabatic, making it crucial to test whether SIDM gravothermal collapse remains robust in the presence of energy injection from baryonic feedback.

In this work, we perform a stress test of SIDM gravothermal collapse under strong baryonic feedback and show that it is resilient to such effects. We implement a semi-analytical feedback model based on an oscillating baryonic potential, calibrated using hydrodynamical CDM simulations and motivated by cores in dwarf galaxies. A key element is the competition between the oscillation timescale of the baryonic potential and the SIDM thermalization timescale. High-concentration halos experience only a mild delay in collapse, which is never halted even under extremely strong feedback. In contrast, median-concentration halos can undergo substantial delays, but collapse resumes once feedback ceases. Their final density profiles depend sensitively on the episodic feedback history, leading to a broader diversity than in CDM. These results further strengthen gravothermal collapse as a smoking-gun signature of SIDM.

\textit{\textbf{Initial dark matter halo--}}We focus on dark matter halos with mass $M_{200}=2\times10^{9}\,\rm{M_{\odot}}$ and two concentrations, $c_{200}=42$ and $c_{200}=15$, corresponding to a $3\sigma$ upward fluctuation and the cosmological median at $z=0$~\cite{Diemer:2018vmz}. We assume that the initial halo follows a Navarro--Frenk--White density profile~\cite{Navarro:1996gj},
\begin{equation}
\rho_{\mathrm{NFW}}(r)=\frac{\rho_s r_s^3}{r\left(r+r_s\right)^2},
\end{equation}
where $\rho_s$ and $r_s$ are the scale density and scale radius, respectively. For the high-concentration halo, we adopt $\rho_{s}=2.32\times10^{8}\,\rm{M_{\odot}/kpc^{3}}$ and $r_{s}=0.63\,\rm{kpc}$, resulting in a maximum circular velocity of $V_{\rm max}\approx33\,\rm{km/s}$. For the median-concentration halo, we take $\rho_{s}=1.74\times10^{7}\,\rm{M_{\odot}/kpc^{3}}$ and $r_{s}=1.69\,\rm{kpc}$ ($V_{\rm max}\approx24\,\rm{km/s}$).

The high-concentration halo is motivated by the JVAS B1938+666 strong-lensing perturber with mass $\sim10^9\,{\rm M_\odot}$~\cite{Tajalli:2025qjx}. Ref.~\cite{Kong:2025sqx} showed that core-collapsed SIDM halos in the Concerto simulation suite~\cite{Nadler:2025jwh} can account for the perturber's high density, while their CDM counterparts typically have concentrations around $c_{200}\approx40$. The median-concentration halo is motivated by the Milky Way satellite Crater II. It is extremely diffuse and has unusually low stellar velocity dispersions~\cite{Torrealba:2016yem,Caldwell:2016hrl,Ji:2021ApJ}, implying a $1\,{\rm kpc}$ dark matter core that is difficult to reproduce in CDM \cite{Borukhovetskaya:2021ahz,Zhang:2024ggu}. In SIDM interpretations of Crater II~\cite{Zhang:2024ggu}, the simulated halo mass is $3\times10^9\,{\rm M_\odot}$ with a concentration close to the cosmological median. Therefore, the two initial halos represent opposite extremes of the density diversity at similar halo mass. Moreover, median-concentration halos are far more common and are representative of typical galaxy host halos.

For the SIDM run, we fix the self-interaction cross section per unit mass to $\sigma/m=50\,{\rm cm^2/g}$, consistent with effective cross sections in velocity-dependent SIDM models in the Concerto suite~\cite{Nadler:2025jwh}, which reproduce the high densities of strong-lensing perturbers~\cite{Nadler:2023nrd,Kong:2025sqx}. This choice is also consistent with SIDM models used to explain Crater II~\cite{Zhang:2024ggu}. For the dark matter halo, we use live particles and conduct controlled N-body simulations with the code \texttt{GADGET-2} \cite{Springel:2000yr, Springel:2005mi}, implemented with an SIDM module~\cite{Yang:2020iya, Yang:2022hkm}. We use the public code \texttt{SpherIC} \cite{Garrison-Kimmel:2013yys} to generate the initial condition for the halo. All simulations included in this work have total $2\times10^{6}$ particles with a Plummer-equivalent softening length $\epsilon=10\,{\rm pc}$ following the suggestion in~\cite{vandenBosch:2018tyt}.

\textit{\textbf{Feedback model--}}Motivated by previous studies~\cite{Pontzen:2011ty,Burger:2021sep}, we adopt a simple tunable model that can generate dark matter cores with controlled sizes. In our controlled N-body simulations, we implement a central Plummer potential~\cite{1911MNRAS..71..460P}, whose effect on the halo is included as an additional acceleration term in the force calculation,
\begin{equation}
\label{eqn:plm-a}
\mathbf{a}_{i, \mathrm{ext}}=-\frac{G M_{\mathrm{ext}}(t)}{\left(a^2+\mathbf{r}_{i}^2\right)^{3 / 2}}\mathbf{r}_i,
\end{equation}
where $\mathbf{r}_{i}$ is the position of the dark matter particle relative to the halo center, $a$ is the Plummer scale radius, and $M_{\rm ext}(t)$ is the time-dependent mass component given by 
\begin{equation}
\label{eqn: m-time-sin}
M_{\rm ext}(t^{\prime}) =
\begin{cases}
M_{\rm max} \sin\!\left(\dfrac{2\pi t^{\prime}}{P}\right), 
   \quad \text{(\textit{\rm {Strong Feedback}})} \\[6pt]
M_{\rm max} \left|\sin\!\left(\dfrac{2\pi t^{\prime}}{P}\right)\right|.
   \quad \text{(\textit{\rm {Weak Feedback}})}
\end{cases}
\end{equation}
$M_{\rm max}$ is the maximum mass and $t^{\prime}$ is the simulation time modulo the feedback period $P$. In the strong-feedback case, $M_{\rm ext}(t^{\prime})$ can become negative, generating a ``repulsive'' acceleration, which drives strong feedback effects. A negative mass is clearly unphysical and should instead be regarded as an effective description of outflows~\cite{Burger:2021sep}.

We fix $M_{\rm max}=10^7\,{\rm M_\odot}$, $a=0.1\,{\rm kpc}$, and $P=0.2\,\rm{Gyr}$, motivated by hydrodynamic simulations~\cite{2023MNRAS.520..461J,2025arXiv250508861P}. We have verified that our strong-feedback model produces core sizes comparable to, though somewhat larger than, those in Ref.~\cite{2023MNRAS.520..461J}. We further vary $P$ and $a$ and find that core growth depends mainly on the total number of feedback cycles, while remaining insensitive to $P$ provided it is shorter than or comparable to the dynamical timescale, consistent with Ref.~\cite{Burger:2021sep}. For compact potentials with $a<0.5\,\rm{kpc}$, the maximum core size scales approximately linearly with $M_{\rm max}$, since the induced acceleration at $r\gg a$ acts as velocity impulses proportional to $M_{\rm max}$. We emphasize that the precise feedback prescription is not essential: even if feedback is extremely strong, gravothermal collapse of high-concentration halos remains robust, as we will show later.

\begin{figure}[t!]
\includegraphics[width=\linewidth]{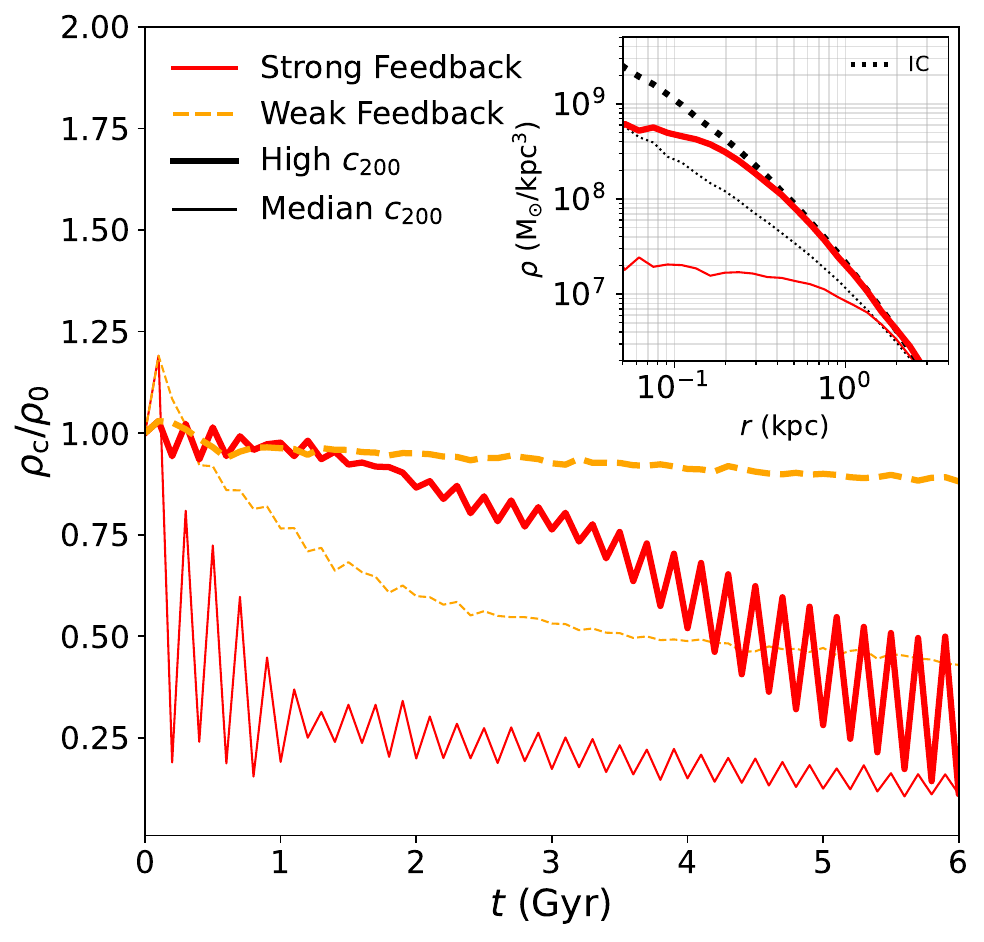}
\caption{\textbf{Main}: Evolution of the normalized inner densities within $0.3\,{\rm kpc}$. The thick (thin) solid red curve represents the high- (median-) concentration halo with the strong feedback model, while the thick (thin) dashed orange curve represents the corresponding halo with the weak feedback model. \textbf{Inset}: Corresponding dark matter density profiles at $t=4\,\rm{Gyr}$ for the strong feedback model, where the halos are relaxed for $1\,\rm{Gyr}$ after the final feedback cycle. The thick (thin) dotted curve denotes the initial high- (median-) concentration halo.}
\label{fig: method}
\end{figure}

\begin{figure*}[t!]
\centering

\includegraphics[width=0.49\textwidth]{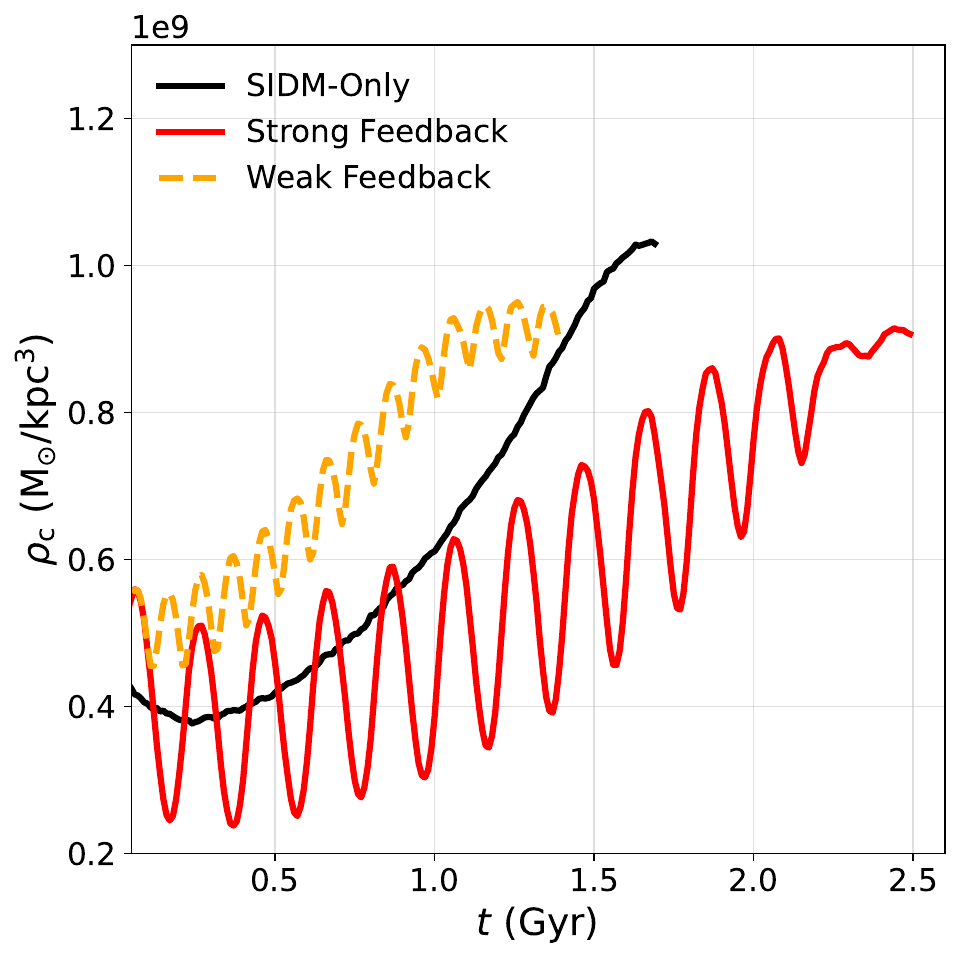}
\includegraphics[width=0.49\textwidth]{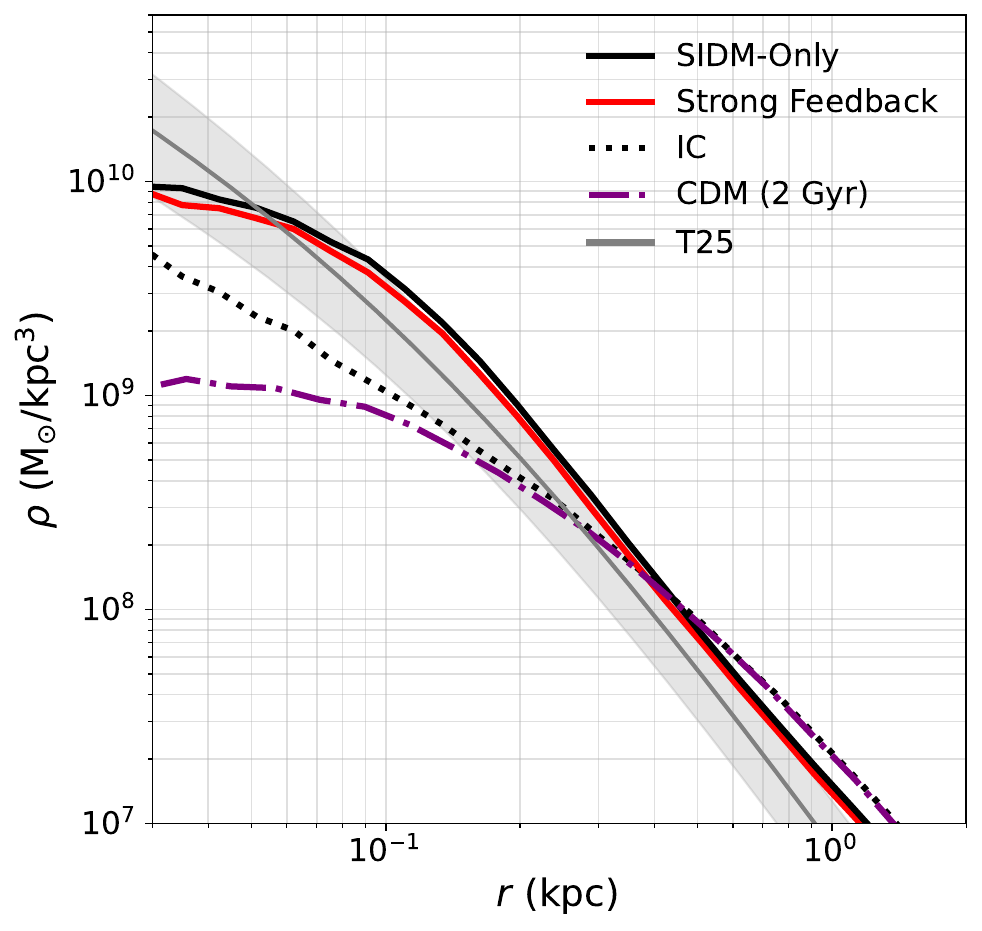}
\caption{\textbf{Left}: Evolution of the averaged inner density within $0.3\,{\rm kpc}$ for the high-concentration halo under different feedback scenarios. The red and orange curves represent the strong (solid) and weak (dashed) feedback models, respectively, while the black curve denotes the SIDM-only simulation. \textbf{Right}: Dark matter density profiles. The solid red curve shows the core-collapsed SIDM halo with the strong feedback model, while the solid black curve denotes the SIDM-only simulation. The dashed black curve represents the initial NFW halo, and the dot-dashed purple curve shows the CDM halo evolved for $2\,\mathrm{Gyr}$ under the oscillating potential of the strong feedback model. The gray curve and shaded region represent the strong-lensing perturber model for the B1938+666 system from Ref.~\cite{Tajalli:2025qjx} (T25).}
\label{fig:haloc}
\end{figure*}

In the main panel of Fig.~\ref{fig: method}, we show the evolution of the central density, averaged within $r=0.3\,\rm{kpc}$ and normalized to the initial value, for high- (thick) and median-concentration (thin) CDM halos under strong (solid red) and weak (dashed orange) feedback models. For strong feedback, both halos exhibit declining central densities with pronounced oscillations. In the high-concentration case, the oscillation amplitude grows with time as repeated energy injection progressively depletes the central region. In contrast, the median-concentration halo rapidly develops a large core within $\sim1\,{\rm Gyr}$, reducing its central density to $\sim25\%$ of the initial value and then remaining nearly constant, with suppressed oscillations due to the low central particle content. For weak feedback, the effect on the high-concentration halo is negligible, while the median-concentration halo is significantly flattened, with an asymptotic reduction of about $50\%$. The inset shows density profiles at $t=4\,{\rm Gyr}$ (solid), compared to the initial conditions (dotted), with core sizes of $0.2\,{\rm kpc}$ and $0.8\,{\rm kpc}$ for the high- and median-concentration halos, respectively.

Under the strong feedback model, the median-concentration halo develops a core size and central density broadly consistent with those inferred for the Milky Way satellite galaxy Crater II~\cite{Caldwell:2016hrl}. In contrast, CDM halos of similar mass in FIRE-2 simulations show the core size is less than ${\cal O}(10)\,{\rm pc}$~\cite{Lazar:2020pjs}, insufficient to explain the $1\,{\rm kpc}$ density core in Crater II. More recent hydrodynamical simulations~\cite{2025arXiv251110582S} indicate that $\sim10^{9}\,\rm{M_{\odot}}$ CDM halos remain cuspy and contain significantly less central gas than assumed here, implying that our strong feedback model represents an extreme limit. Nevertheless, gravothermal collapse in SIDM remains robust even under such extreme feedback, as we show next.

\textit{\textbf{High-concentration halo--}} In the left panel of Fig.~\ref{fig:haloc}, we present the central density evolution for the high-concentration SIDM halo under strong (red) and weak (orange) feedback, together with the SIDM-only case (black). For each case shown in Fig.~\ref{fig:haloc}, we truncate the evolution before core collapse becomes artificially stalled by numerical heating~\cite{Palubski:2024ibb,Fischer:2024eaz,Fischer:2025rky}. In the SIDM-only run, the central density reaches a minimum at $t\approx0.3\,\mathrm{Gyr}$, after which the halo enters the collapse phase and the central density increases continuously and smoothly.

With the strong feedback model, the central density fluctuates by a factor of $\sim2$ per cycle, with the amplitude gradually decreasing as the halo deepens into collapse. Nevertheless, core collapse is {\it not} prevented, and reaching SIDM-only densities is delayed by only $\sim1\,\mathrm{Gyr}$. In contrast, the weak feedback model induces smaller fluctuations and mildly accelerates collapse due to contraction during the inflow phase, consistent with earlier studies of static baryonic potentials~\cite{Feng:2020kxv,Zhong:2023yzk,Yang:2024tba,Jia:2026ocr}.

In the right panel of Fig.~\ref{fig:haloc}, we show the SIDM density profiles under strong feedback (solid red) and SIDM-only (solid black) at their final snapshot. For comparison, we include the strong-lensing perturber model for the B1938+666 system~\cite{Tajalli:2025qjx} (T25; shaded gray), along with the initial NFW halo (dashed black) and the CDM halo evolved under strong feedback ($t=2\,{\rm Gyr}$, dot-dashed purple). Both SIDM cases remain consistent with the inferred perturber constraints, while the initial NFW profile is only marginally consistent at $2\sigma$. In contrast, the CDM halo with feedback becomes too shallow due to core formation, failing to match the perturber model. 

For the high-concentration halo, the thermalization timescale is $t_{\rm th}\sim0.01\textup{--}0.04\,{\rm Gyr}$ throughout the evolution, much shorter than the oscillation period $P=0.2\,{\rm Gyr}$ (see End Matter). Thus, SIDM dynamics dominate over baryonic feedback, and gravothermal collapse remains robust even under extreme feedback. Since the core-collapse timescale is highly sensitive to the concentration $t_{\rm cc}\propto c_{200}^{-7/2}$~\cite{Essig:2018pzq}, high-concentration halos are prime targets for probing this characteristic SIDM prediction. Our results further strengthen SIDM core-collapse interpretations of the dense strong-lensing perturbers~\cite{Nadler:2023nrd,Kong:2025sqx,Li:2025kpb,Yu:2025tmp}, with broader implications for further observational tests~\cite{Gilman:2022ida,Meneghetti:2020yif,Dutra:2024qac,Hou:2025gmv,Natarajan:2026ztd,Kollmann:2025czq,Mace:2026gxg}. SIDM core collapse may also provide a mechanism for seeding black holes associated with JWST little red dots \cite{Jiang:2025jtr,Feng:2025rzf}. In this scenario, early-forming high-concentration halos could host a significant population of seed black holes, and our results suggest that the collapse remains robust even that star formation is highly bursty in the early Universe~\cite{Clarke:2025}.

\begin{figure}[t]
\includegraphics[scale=0.49]{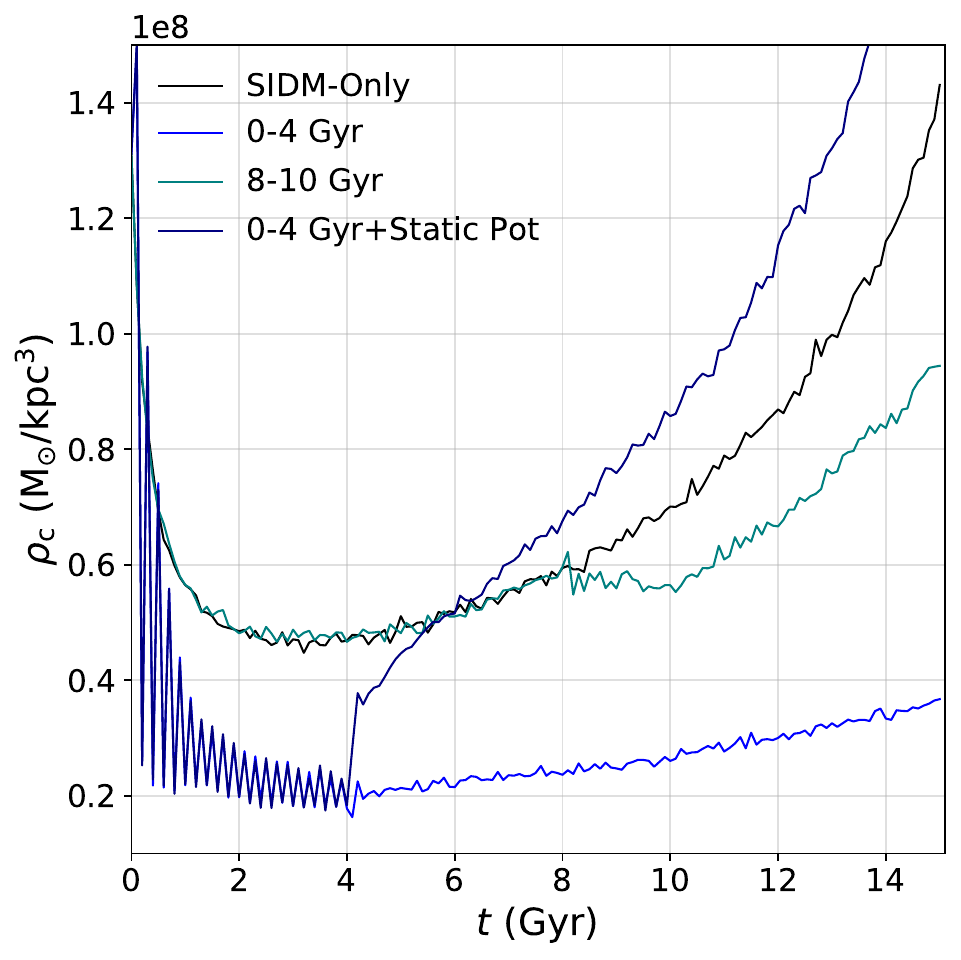}
\caption{Central density evolution of the median-concentration halo for different feedback scenarios. The blue curve denotes the case in which feedback is active during $0\textup{--}4\,\rm{Gyr}$ and completely removed afterward. The navy curve denotes the case in which feedback is active during $0\textup{--}4\,\rm{Gyr}$ and subsequently replaced by a static baryonic potential with $M=M_{\rm max}$ for the remainder of the evolution. The teal curve represents the case in which feedback is active during $8\textup{--}10\,\rm{Gyr}$. The black curve shows the SIDM-only case for comparison.
}
\label{fig:period}
\end{figure}

\textit{\textbf{Median-concentration halo--}}We next investigate the episodic impacts of baryonic feedback on gravothermal collapse. We focus on the median-concentration halo, as its relatively long core-collapse timescale makes it more susceptible to baryonic feedback during gravothermal evolution. To explore the maximal effects, we apply the strong feedback model.

Figure~\ref{fig:period} shows the central density evolution of the median-concentration halo under different feedback histories. For feedback active during $0\textup{--}4\,{\rm Gyr}$ and then removed (blue), the central density is reduced by a factor of $\sim2.5$ relative to the SIDM-only case (black), reflecting the dominant effect of the oscillating potential during the core-expansion phase. After feedback ceases at $t=4\,{\rm Gyr}$, gravothermal collapse resumes, albeit more slowly, demonstrating that strong feedback cannot permanently prevent collapse. For this halo, the thermalization timescale is longer than the oscillation period for most of its evolution (see End Matter).

When the oscillating potential is replaced at $t=4\,{\rm Gyr}$ by a static maximum-strength potential (navy), collapse is instead accelerated, reaching the SIDM-only central density $2\,{\rm Gyr}$ earlier. This shows that the residual baryonic potential after star formation can largely determine the final evolution. For late-time feedback during $8\textup{--}10\,{\rm Gyr}$ (teal), the halo is already in the collapse phase; feedback temporarily halts but does not reverse it, and collapse resumes once feedback ends. Overall, median-concentration halos are highly sensitive to episodic feedback: it can delay or modulate collapse, leading to a wide diversity of final density profiles compared to CDM.

It would be interesting to explore whether the combined effects of self-interactions and baryonic feedback can explain the properties of gas-rich ultra-diffuse galaxies reported in \cite{ManceraPina:2019zih,ManceraPina:2020ujo,PinaMancera:2021wpc}, which are expected to reside in low-concentration halos. Such systems are difficult to reproduce in either feedback-only or SIDM-only scenarios~\cite{ManceraPina:2024ybj,Kong:2022oyk}, since low-concentration halos do not efficiently develop large density cores, while repeated feedback cycles alone tend to deplete gas. As shown in Fig.~\ref{fig:period}, if feedback occurs during the early stages of galaxy formation, the combined effects may rapidly generate large dark matter cores while retaining most of the gas.

\textit{\textbf{Conclusion--}}We have investigated the impact of baryonic feedback on gravothermal collapse in SIDM halos using a calibrated semi-analytical model based on an oscillating external potential that captures rapid potential fluctuations from violent feedback. For high-concentration halos, the short thermalization timescale ensures that gravothermal collapse remains robust and only mildly delayed even under strong feedback. For median-concentration halos, feedback during the core-expansion phase can rapidly produce large, shallow cores, but collapse resumes once feedback ceases. We further showed that the final density profiles of such halos depend sensitively on the episodic history of baryonic feedback, producing a broad diversity in central densities.

These results reinforce SIDM core-collapse interpretations of dense compact objects and support SIDM-based scenarios for early supermassive black hole seeding. The predicted diversity of density profiles can be further tested with galaxy rotation curves. Since the semi-analytical feedback model is highly tunable, it can be systematically explored and calibrated against observed galaxy populations. Moreover, it can be incorporated into cosmological simulations in a hybrid scheme. We leave these promising directions for future work.

\begin{acknowledgments}
This work is supported by the U.S. Department of Energy under grant No.~DE-SC0008541 and the John Templeton Foundation under grant No.~63599. Computations were performed using the computer clusters and data storage resources of the UCR HPCC, which were funded by grants from NSF (MRI-2215705, MRI-1429826) and NIH (1S10OD016290-01A1). The opinions expressed in this publication are those of the authors and do not necessarily reflect the views of the funding agencies.
\end{acknowledgments}

\bibliography{apssamp}

@article{Spergel:1999mh,
    author = "Spergel, David N. and Steinhardt, Paul J.",
    title = "{Observational evidence for selfinteracting cold dark matter}",
    eprint = "astro-ph/9909386",
    archivePrefix = "arXiv",
    doi = "10.1103/PhysRevLett.84.3760",
    journal = "Phys. Rev. Lett.",
    volume = "84",
    pages = "3760--3763",
    year = "2000"
}

@article{Shen:2025evo,
    author = "Shen, Tingwei and Shen, Xuejian and Xiao, Huangyu and Vogelsberger, Mark and Jiang, Fangzhou",
    title = "{Massive Black Holes Seeded by Dark Matter -- Implications for Little Red Dots and Gravitational Wave Signatures}",
    eprint = "2504.00075",
    archivePrefix = "arXiv",
    primaryClass = "astro-ph.GA",
    reportNumber = "FERMILAB-PUB-25-0224-T",
    journal = "arXiv e-prints",
    pages = "arXiv:2504.00075",
    month = "3",
    year = "2025"
}

@article{Feng:2021rst,
    author = "Feng, Wei-Xiang and Yu, Hai-Bo and Zhong, Yi-Ming",
    title = "{Dynamical instability of collapsed dark matter halos}",
    eprint = "2108.11967",
    archivePrefix = "arXiv",
    primaryClass = "astro-ph.CO",
    doi = "10.1088/1475-7516/2022/05/036",
    journal = "JCAP",
    volume = "05",
    number = "05",
    pages = "036",
    year = "2022"
}

@article{Gad-Nasr:2023gvf,
    author = "Gad-Nasr, Sophia and Boddy, Kimberly K. and Kaplinghat, Manoj and Outmezguine, Nadav Joseph and Sagunski, Laura",
    title = "{On the late-time evolution of velocity-dependent self-interacting dark matter halos}",
    eprint = "2312.09296",
    archivePrefix = "arXiv",
    primaryClass = "astro-ph.GA",
    doi = "10.1088/1475-7516/2024/05/131",
    journal = "JCAP",
    volume = "05",
    number = "05",
    pages = "131",
    year = "2024"
}

@article{Gu:2026zzq,
    author = "Gu, Hua-Peng and Jiang, Fangzhou and Chen, Xian and Li, Ran",
    title = "{Nonequilibrium relativistic core collapse of self-interacting dark matter halos: Limits on the seed black hole mass}",
    eprint = "2601.17117",
    archivePrefix = "arXiv",
    primaryClass = "astro-ph.CO",
    doi = "10.1103/6sl3-kjzf",
    journal = "Phys. Rev. D",
    volume = "113",
    number = "10",
    pages = "103038",
    year = "2026"
}

@article{KuziodeNaray:2009oon,
    author = "Kuzio de Naray, Rachel and Martinez, Gregory D. and Bullock, James S. and Kaplinghat, Manoj",
    title = "{The Case Against Warm or Self-Interacting Dark Matter as Explanations for Cores in Low Surface Brightness Galaxies}",
    eprint = "0912.3518",
    archivePrefix = "arXiv",
    primaryClass = "astro-ph.CO",
    doi = "10.1088/2041-8205/710/2/L161",
    journal = "Astrophys. J. Lett.",
    volume = "710",
    pages = "L161",
    year = "2010"
}

@article{Cruz:2025wfz,
    author = "Cruz, Akaxia and Brooks, Alyson and Lisanti, Mariangela and Peter, Annika H. G. and Geda, Robel and Quinn, Thomas and Tremmel, Michael and Munshi, Ferah and Keller, Ben and Wadsley, James",
    title = "{Dwarf diversity in $Λ$CDM with baryons}",
    eprint = "2510.11800",
    archivePrefix = "arXiv",
    primaryClass = "astro-ph.GA",
    journal = "arXiv e-prints",
    pages = "arXiv:2510.11800",
    month = "10",
    year = "2025"
}

@article{Yang:2024tba,
    author = "Yang, Daneng",
    title = "{Exploring self-interacting dark matter halos with diverse baryonic distributions: A parametric approach}",
    eprint = "2405.03787",
    archivePrefix = "arXiv",
    primaryClass = "astro-ph.CO",
    doi = "10.1103/PhysRevD.110.103044",
    journal = "Phys. Rev. D",
    volume = "110",
    number = "10",
    pages = "103044",
    year = "2024"
}

@article{Hou:2025gmv,
    author = "Hou, Siyuan and Yang, Daneng and Li, Nan and Li, Guoliang",
    title = "{A universal analytic model for gravitational lensing by self-interacting dark matter halos}",
    eprint = "2502.14964",
    archivePrefix = "arXiv",
    primaryClass = "astro-ph.CO",
    doi = "10.1088/1475-7516/2025/08/048",
    journal = "JCAP",
    volume = "08",
    number = "08",
    pages = "048",
    year = "2025"
}

@article{Despali:2025koj,
    author = "Despali, Giulia and Moscardini, Lauro and Nelson, Dylan and Pillepich, Annalisa and Springel, Volker and Vogelsberger, Mark",
    title = "{Introducing the AIDA-TNG project: Galaxy formation in alternative dark matter models}",
    eprint = "2501.12439",
    archivePrefix = "arXiv",
    primaryClass = "astro-ph.CO",
    doi = "10.1051/0004-6361/202553836",
    journal = "Astron. Astrophys.",
    volume = "697",
    pages = "A213",
    year = "2025"
}

@article{Engelhardt:2026dpj,
    author = "Engelhardt, Anna and Munshi, Ferah and Peter, Annika H. G. and Nadler, Ethan O. and Cruz, Akaxia and Brooks, Alyson M. and Zeng, Zhichao Carton and Quinn, Thomas R. and Keith, Blake",
    title = "{MARVELously Dark: the gravothermal evolution of dwarf halos in velocity-dependent SIDM}",
    eprint = "2601.23264",
    archivePrefix = "arXiv",
    primaryClass = "astro-ph.GA",
    journal = "arXiv e-prints",
    pages = "arXiv:2601.23264",
    month = "1",
    year = "2026"
}

@article{Fischer:2023lvl,
    author = {Fischer, Moritz S. and Kasselmann, Lenard and Br{\"u}ggen, Marcus and Dolag, Klaus and Kahlhoefer, Felix and Ragagnin, Antonio and Robertson, Andrew and Schmidt-Hoberg, Kai},
    title = "{Cosmological and idealized simulations of dark matter haloes with velocity-dependent, rare and frequent self-interactions}",
    eprint = "2310.07750",
    archivePrefix = "arXiv",
    primaryClass = "astro-ph.CO",
    doi = "10.1093/mnras/stae699",
    journal = "Mon. Not. Roy. Astron. Soc.",
    volume = "529",
    number = "3",
    pages = "2327--2348",
    year = "2024"
}

@article{Yang:2022mxl,
    author = "Yang, Daneng and Nadler, Ethan O. and Yu, Hai-Bo",
    title = "{Strong Dark Matter Self-interactions Diversify Halo Populations within and surrounding the Milky Way}",
    eprint = "2211.13768",
    archivePrefix = "arXiv",
    primaryClass = "astro-ph.GA",
    doi = "10.3847/1538-4357/acc73e",
    journal = "Astrophys. J.",
    volume = "949",
    number = "2",
    pages = "67",
    year = "2023"
}

@article{Correa:2022dey,
    author = "Correa, Camila A. and Schaller, Matthieu and Ploeckinger, Sylvia and Anau Montel, Noemi and Weniger, Christoph and Ando, Shin{\textquoteright}ichiro",
    title = "{TangoSIDM: tantalizing models of self-interacting dark matter}",
    eprint = "2206.11298",
    archivePrefix = "arXiv",
    primaryClass = "astro-ph.GA",
    doi = "10.1093/mnras/stac2830",
    journal = "Mon. Not. Roy. Astron. Soc.",
    volume = "517",
    number = "2",
    pages = "3045--3063",
    year = "2022"
}

@article{Zavala:2019sjk,
    author = "Zavala, Jes{\'u}s and Lovell, Mark R. and Vogelsberger, Mark and Burger, Jan D.",
    title = "{Diverse dark matter density at sub-kiloparsec scales in Milky Way satellites: Implications for the nature of dark matter}",
    eprint = "1904.09998",
    archivePrefix = "arXiv",
    primaryClass = "astro-ph.GA",
    doi = "10.1103/PhysRevD.100.063007",
    journal = "Phys. Rev. D",
    volume = "100",
    number = "6",
    pages = "063007",
    year = "2019"
}

@article{Slone:2021nqd,
    author = "Slone, Oren and Jiang, Fangzhou and Lisanti, Mariangela and Kaplinghat, Manoj",
    title = "{Orbital evolution of satellite galaxies in self-interacting dark matter models}",
    eprint = "2108.03243",
    archivePrefix = "arXiv",
    primaryClass = "astro-ph.CO",
    doi = "10.1103/PhysRevD.107.043014",
    journal = "Phys. Rev. D",
    volume = "107",
    number = "4",
    pages = "043014",
    year = "2023"
}

@article{Correa:2020qam,
    author = "Correa, Camila A.",
    title = "{Constraining velocity-dependent self-interacting dark matter with the Milky Way{\textquoteright}s dwarf spheroidal galaxies}",
    eprint = "2007.02958",
    archivePrefix = "arXiv",
    primaryClass = "astro-ph.GA",
    doi = "10.1093/mnras/stab506",
    journal = "Mon. Not. Roy. Astron. Soc.",
    volume = "503",
    number = "1",
    pages = "920--937",
    year = "2021"
}

@article{Sameie:2019zfo,
    author = "Sameie, Omid and Yu, Hai-Bo and Sales, Laura V. and Vogelsberger, Mark and Zavala, Jes{\'u}s",
    title = "{Self-Interacting Dark Matter Subhalos in the Milky Way{\textquoteright}s Tides}",
    eprint = "1904.07872",
    archivePrefix = "arXiv",
    primaryClass = "astro-ph.GA",
    doi = "10.1103/PhysRevLett.124.141102",
    journal = "Phys. Rev. Lett.",
    volume = "124",
    number = "14",
    pages = "141102",
    year = "2020"
}

@ARTICLE{Clarke:2025,
       author = {{Clarke}, Leonardo and {Shapley}, Alice E. and {Lam}, Natalie and {Topping}, Michael W. and {Brammer}, Gabriel B. and {Sanders}, Ryan L. and {Reddy}, Naveen A. and {Karthikeyan}, Shreya},
        title = "{The Star-forming Main Sequence and Bursty Star-formation Histories at $z>1.4$ in JADES and AURORA}",
      journal = {arXiv e-prints},
         year = 2025,
        month = oct,
          eid = {arXiv:2510.06681},
        pages = {arXiv:2510.06681},
          doi = {10.48550/arXiv.2510.06681},
archivePrefix = {arXiv},
       eprint = {2510.06681},
 primaryClass = {astro-ph.GA},
       adsurl = {https://ui.adsabs.harvard.edu/abs/2025arXiv251006681C}
}

@article{Natarajan:2026ztd,
    author = "Natarajan, Priyamvada and Chiang, Barry T. and Dutra, Isaque",
    title = "{New Cold Dark Matter Crisis Revealed by Multiscale Cluster Lensing}",
    eprint = "2601.07909",
    archivePrefix = "arXiv",
    primaryClass = "astro-ph.CO",
    doi = "10.3847/2041-8213/ae53ea",
    journal = "Astrophys. J. Lett.",
    volume = "1001",
    number = "1",
    pages = "L12",
    year = "2026"
}

@article{Meneghetti:2020yif,
    author = "Meneghetti, Massimo and others",
    title = "{An excess of small-scale gravitational lenses observed in galaxy clusters}",
    eprint = "2009.04471",
    archivePrefix = "arXiv",
    primaryClass = "astro-ph.GA",
    doi = "10.1126/science.aax5164",
    journal = "Science",
    volume = "369",
    number = "6509",
    pages = "1347--1351",
    year = "2020"
}

@article{Dutra:2024qac,
    author = "Dutra, Isaque and Natarajan, Priyamvada and Gilman, Daniel",
    title = "{Self-interacting Dark Matter, Core Collapse, and the Galaxy{\textendash}Galaxy Strong-lensing Discrepancy}",
    eprint = "2406.17024",
    archivePrefix = "arXiv",
    primaryClass = "astro-ph.CO",
    doi = "10.3847/1538-4357/ad9b09",
    journal = "Astrophys. J.",
    volume = "978",
    number = "1",
    pages = "38",
    year = "2025"
}

@article{Mace:2026gxg,
    author = "Mace, Charlie and Dhanasingham, Birendra and Zeng, Zhichao Carton and Cyr-Racine, Francis-Yan and Du, Xiaolong and Peter, Annika H. G. and Benson, Andrew",
    title = "{The Sensitivity of Substructure Lensing to SIDM Core-collapse Model Variation}",
    eprint = "2605.24174",
    archivePrefix = "arXiv",
    primaryClass = "astro-ph.CO",
    journal = "arXiv e-prints",
    pages = "arXiv:2605.24174",
    month = "5",
    year = "2026"
}

@article{Kollmann:2025czq,
    author = "Kollmann, Kassidy E. and Nightingale, James W. and Lisanti, Mariangela and Robertson, Andrew and Slone, Oren",
    title = "{Using strong lensing to detect subhaloes with steep inner density profiles}",
    eprint = "2510.17956",
    archivePrefix = "arXiv",
    primaryClass = "astro-ph.CO",
    doi = "10.1093/mnras/stag066",
    journal = "Mon. Not. Roy. Astron. Soc.",
    volume = "546",
    number = "4",
    pages = "stag066",
    year = "2026"
}

@article{Gilman:2022ida,
    author = "Gilman, Daniel and Zhong, Yi-Ming and Bovy, Jo",
    title = "{Constraining resonant dark matter self-interactions with strong gravitational lenses}",
    eprint = "2207.13111",
    archivePrefix = "arXiv",
    primaryClass = "astro-ph.CO",
    doi = "10.1103/PhysRevD.107.103008",
    journal = "Phys. Rev. D",
    volume = "107",
    number = "10",
    pages = "103008",
    year = "2023"
}

@ARTICLE{Ji:2021ApJ,
       author = {{Ji}, Alexander P. and {Koposov}, Sergey E. and {Li}, Ting S. and {Erkal}, Denis and {Pace}, Andrew B. and {Simon}, Joshua D. and {Belokurov}, Vasily and {Cullinane}, Lara R. and {Da Costa}, Gary S. and {Kuehn}, Kyler and {Lewis}, Geraint F. and {Mackey}, Dougal and {Shipp}, Nora and {Simpson}, Jeffrey D. and {Zucker}, Daniel B. and {Hansen}, Terese T. and {Bland-Hawthorn}, Joss and {S5 Collaboration}},
        title = "{Kinematics of Antlia 2 and Crater 2 from the Southern Stellar Stream Spectroscopic Survey (S$^{5}$)}",
      journal = {\apj},
         year = 2021,
        month = nov,
       volume = {921},
       number = {1},
          eid = {32},
        pages = {32},
          doi = {10.3847/1538-4357/ac1869},
archivePrefix = {arXiv},
       eprint = {2106.12656},
 primaryClass = {astro-ph.GA},
       adsurl = {https://ui.adsabs.harvard.edu/abs/2021ApJ...921...32J}
}

@article{Torrealba:2016yem,
    author = "Torrealba, G. and Koposov, S. E. and Belokurov, V. and Irwin, M.",
    title = "{The feeble giant. Discovery of a large and diffuse Milky Way dwarf galaxy in the constellation of Crater}",
    eprint = "1601.07178",
    archivePrefix = "arXiv",
    doi = "10.1093/mnras/stw733",
    journal = "Mon. Not. Roy. Astron. Soc.",
    volume = "459",
    number = "3",
    pages = "2370--2378",
    year = "2016"
}

@article{Vegetti:2026mmx,
    author = "Vegetti, Simona and White, Simon D. M. and McKean, John P. and Powell, Devon M. and Spingola, Cristiana and Massari, Davide and Despali, Giulia and Fassnacht, Christopher D.",
    title = "{A possible challenge for cold and warm dark matter}",
    eprint = "2601.02466",
    archivePrefix = "arXiv",
    primaryClass = "astro-ph.CO",
    doi = "10.1038/s41550-025-02746-w",
    journal = "Nature Astron.",
    volume = "10",
    number = "3",
    pages = "440--447",
    year = "2026"
}

@article{Fischer:2026ryr,
    author = "Fischer, Moritz S. and Yu, Hai-Bo",
    title = "{The dark fate of ultra-faint dwarfs: gravothermal collapse in action}",
    eprint = "2603.04508",
    archivePrefix = "arXiv",
    primaryClass = "astro-ph.CO",
    journal = "arXiv e-prints",
    pages = "arXiv:2603.04508",
    month = "3",
    year = "2026"
}

@article{Sengul:2021lxe,
    author = {{\c{S}}eng{\"u}l, At{\i}n{\c{c}} {\c{C}}a{\u{g}}an and Dvorkin, Cora and Ostdiek, Bryan and Tsang, Arthur},
    title = "{Substructure detection reanalysed: dark perturber shown to be a line-of-sight halo}",
    eprint = "2112.00749",
    archivePrefix = "arXiv",
    primaryClass = "astro-ph.CO",
    doi = "10.1093/mnras/stac1967",
    journal = "Mon. Not. Roy. Astron. Soc.",
    volume = "515",
    number = "3",
    pages = "4391--4401",
    year = "2022"
}

@article{Enzi:2024ygw,
    author = "Enzi, Wolfgang J. R. and Krawczyk, Coleman M. and Ballard, Daniel J. and Collett, Thomas E.",
    title = "{The overconcentrated dark halo in the strong lens SDSS J0946~+~1006 is a subhalo: evidence for self-interacting dark matter?}",
    eprint = "2411.08565",
    archivePrefix = "arXiv",
    primaryClass = "astro-ph.CO",
    doi = "10.1093/mnras/staf697",
    journal = "Mon. Not. Roy. Astron. Soc.",
    volume = "540",
    number = "1",
    pages = "247--263",
    year = "2025"
}

@ARTICLE{Vegetti:2010,
       author = {{Vegetti}, S. and {Koopmans}, L.~V.~E. and {Bolton}, A. and {Treu}, T. and {Gavazzi}, R.},
        title = "{Detection of a dark substructure through gravitational imaging}",
      journal = {\mnras},
         year = 2010,
        month = nov,
       volume = {408},
       number = {4},
        pages = {1969-1981},
          doi = {10.1111/j.1365-2966.2010.16865.x},
archivePrefix = {arXiv},
       eprint = {0910.0760},
 primaryClass = {astro-ph.CO},
       adsurl = {https://ui.adsabs.harvard.edu/abs/2010MNRAS.408.1969V}
}

@article{Powell:2025rmj,
    author = "Powell, D. M. and McKean, J. P. and Vegetti, S. and Spingola, C. and White, S. D. M. and Fassnacht, C. D.",
    title = "{A million-solar-mass object detected at a cosmological distance using gravitational imaging}",
    eprint = "2510.07382",
    archivePrefix = "arXiv",
    primaryClass = "astro-ph.CO",
    doi = "10.1038/s41550-025-02651-2",
    journal = "Nature Astron.",
    volume = "9",
    number = "11",
    pages = "1714--1722",
    year = "2025"
}

@article{Jia:2026ocr,
    author = "Jia, Zixiang and Jiang, Fangzhou and Li, Shubo and Li, Ran and Wang, Jing and Zhu, Ling",
    title = "{An Enhanced Isothermal Jeans Approach to Constraining Dark Matter Self-Interactions from Galactic Kinematics}",
    eprint = "2601.17118",
    archivePrefix = "arXiv",
    primaryClass = "astro-ph.GA",
    journal = "arXiv e-prints",
    pages = "arXiv:2601.17118",
    month = "1",
    year = "2026"
}

@article{Navarro:1996bv,
    author = "Navarro, Julio F. and Eke, Vincent R. and Frenk, Carlos S.",
    title = "{The cores of dwarf galaxy halos}",
    eprint = "astro-ph/9610187",
    archivePrefix = "arXiv",
    doi = "10.1093/mnras/283.3.L72",
    journal = "Mon. Not. Roy. Astron. Soc.",
    volume = "283",
    pages = "L72--L78",
    year = "1996"
}

@article{Oh:2015xoa,
    author = "Oh, Se-Heon and others",
    title = "{High-resolution mass models of dwarf galaxies from LITTLE THINGS}",
    eprint = "1502.01281",
    archivePrefix = "arXiv",
    primaryClass = "astro-ph.GA",
    doi = "10.1088/0004-6256/149/6/180",
    journal = "Astron. J.",
    volume = "149",
    pages = "180",
    year = "2015"
}

@article{Sales:2022ich,
    author = "Sales, Laura V. and Wetzel, Andrew and Fattahi, Azadeh",
    title = "{Baryonic solutions and challenges for cosmological models of dwarf galaxies}",
    eprint = "2206.05295",
    archivePrefix = "arXiv",
    primaryClass = "astro-ph.GA",
    doi = "10.1038/s41550-022-01689-w",
    journal = "Nature Astron.",
    volume = "6",
    number = "8",
    pages = "897--910",
    year = "2022"
}

@article{Gentile:2004tb,
    author = "Gentile, Gianfranco and Salucci, P. and Klein, U. and Vergani, D. and Kalberla, P.",
    title = "{The Cored distribution of dark matter in spiral galaxies}",
    eprint = "astro-ph/0403154",
    archivePrefix = "arXiv",
    doi = "10.1111/j.1365-2966.2004.07836.x",
    journal = "Mon. Not. Roy. Astron. Soc.",
    volume = "351",
    pages = "903",
    year = "2004"
}

@article{Kong:2022oyk,
    author = "Kong, Demao and Kaplinghat, Manoj and Yu, Hai-Bo and Fraternali, Filippo and Pi{\~n}a Mancera, Pavel E.",
    title = "{The Odd Dark Matter Halos of Isolated Gas-rich Ultradiffuse Galaxies}",
    eprint = "2204.05981",
    archivePrefix = "arXiv",
    primaryClass = "astro-ph.GA",
    doi = "10.3847/1538-4357/ac8875",
    journal = "Astrophys. J.",
    volume = "936",
    number = "2",
    pages = "166",
    year = "2022"
}

@article{ManceraPina:2020ujo,
    author = "Mancera Pi{\~n}a, Pavel E. and others",
    title = "{Robust HI kinematics of gas-rich ultra-diffuse galaxies: hints of a weak-feedback formation scenario}",
    eprint = "2004.14392",
    archivePrefix = "arXiv",
    primaryClass = "astro-ph.GA",
    doi = "10.1093/mnras/staa1256",
    journal = "Mon. Not. Roy. Astron. Soc.",
    volume = "495",
    number = "4",
    pages = "3636--3655",
    year = "2020"
}

@article{PinaMancera:2021wpc,
    author = "Pi{\~n}a Mancera, Pavel E. and Fraternali, Filippo and Oosterloo, Tom and Adams, Elizabeth A. K. and Oman, Kyle A. and Leisman, Lukas",
    title = "{No need for dark matter: resolved kinematics of the ultra-diffuse galaxy AGC 114905}",
    eprint = "2112.00017",
    archivePrefix = "arXiv",
    primaryClass = "astro-ph.GA",
    doi = "10.1093/mnras/stab3491",
    journal = "Mon. Not. Roy. Astron. Soc.",
    volume = "512",
    number = "3",
    pages = "3230--3242",
    year = "2022"
}

@article{ManceraPina:2024ybj,
    author = "Mancera Pi{\~n}a, Pavel E. and Golini, Giulia and Trujillo, Ignacio and Montes, Mireia",
    title = "{Exploring the nature of dark matter with the extreme galaxy AGC 114905}",
    eprint = "2404.06537",
    archivePrefix = "arXiv",
    primaryClass = "astro-ph.GA",
    doi = "10.1051/0004-6361/202450230",
    journal = "Astron. Astrophys.",
    volume = "689",
    pages = "A344",
    year = "2024"
}

@article{Roberts:2025poo,
    author = "Roberts, M. Grant and Braff, Lila and Garg, Aarna and Profumo, Stefano and Jeltema, Tesla",
    title = "{Little Red Dots from ultra-strongly self-interacting dark matter}",
    eprint = "2507.03230",
    archivePrefix = "arXiv",
    primaryClass = "astro-ph.GA",
    doi = "10.1088/1475-7516/2026/05/003",
    journal = "JCAP",
    volume = "05",
    number = "05",
    pages = "003",
    year = "2026"
}

@article{Feng:2020kxv,
    author = "Feng, Wei-Xiang and Yu, Hai-Bo and Zhong, Yi-Ming",
    title = "{Seeding Supermassive Black Holes with Self-interacting Dark Matter: A Unified Scenario with Baryons}",
    eprint = "2010.15132",
    archivePrefix = "arXiv",
    primaryClass = "astro-ph.CO",
    doi = "10.3847/2041-8213/ac04b0",
    journal = "Astrophys. J. Lett.",
    volume = "914",
    number = "2",
    pages = "L26",
    year = "2021"
}

@article{Zeng:2021ldo,
    author = "Zeng, Zhichao Carton and Peter, Annika H. G. and Du, Xiaolong and Benson, Andrew and Kim, Stacy and Jiang, Fangzhou and Cyr-Racine, Francis-Yan and Vogelsberger, Mark",
    title = "{Core-collapse, evaporation, and tidal effects: the life story of a self-interacting dark matter subhalo}",
    eprint = "2110.00259",
    archivePrefix = "arXiv",
    primaryClass = "astro-ph.CO",
    doi = "10.1093/mnras/stac1094",
    journal = "Mon. Not. Roy. Astron. Soc.",
    volume = "513",
    number = "4",
    pages = "4845--4868",
    year = "2022"
}

@article{Adhikari:2022sbh,
    author = "Adhikari, Susmita and others",
    title = "{Astrophysical tests of dark matter self-interactions}",
    eprint = "2207.10638",
    archivePrefix = "arXiv",
    primaryClass = "astro-ph.CO",
    doi = "10.1103/m2vm-59y3",
    journal = "Rev. Mod. Phys.",
    volume = "97",
    number = "4",
    pages = "045004",
    year = "2025"
}

@article{Despali:2024ihn,
    author = "Despali, Giulia and Heinze, Felix M. and Fassnacht, Christopher D. and Vegetti, Simona and Spingola, Cristiana and Klessen, Ralf and Tajalli, Maryam",
    title = "{Detecting low-mass haloes with strong gravitational lensing - II. Constraints on the density profiles of two detected subhaloes}",
    eprint = "2407.12910",
    archivePrefix = "arXiv",
    primaryClass = "astro-ph.CO",
    doi = "10.1051/0004-6361/202451546",
    journal = "Astron. Astrophys.",
    volume = "699",
    pages = "A222",
    year = "2025"
}

@article{Vegetti:2012mc,
    author = "Vegetti, S. and Lagattuta, D. J. and McKean, J. P. and Auger, M. W. and Fassnacht, C. D. and Koopmans, L. V. E.",
    title = "{Gravitational detection of a low-mass dark satellite at cosmological distance}",
    eprint = "1201.3643",
    archivePrefix = "arXiv",
    primaryClass = "astro-ph.CO",
    doi = "10.1038/nature10669",
    journal = "Nature",
    volume = "481",
    pages = "341",
    year = "2012"
}

@article{Nadler:2025jwh,
    author = "Nadler, Ethan O. and Kong, Demao and Yang, Daneng and Yu, Hai-Bo",
    title = "{SIDM Concerto: Compilation and Data Release of Self-interacting Dark Matter Zoom-in Simulations}",
    eprint = "2503.10748",
    archivePrefix = "arXiv",
    primaryClass = "astro-ph.CO",
    doi = "10.3847/1538-4357/adf553",
    journal = "Astrophys. J.",
    volume = "991",
    number = "1",
    pages = "69",
    year = "2025"
}

@article{Springel:2000yr,
    author = "Springel, Volker and Yoshida, Naoki and White, Simon D. M.",
    title = "{GADGET: A Code for collisionless and gasdynamical cosmological simulations}",
    eprint = "astro-ph/0003162",
    archivePrefix = "arXiv",
    doi = "10.1016/S1384-1076(01)00042-2",
    journal = "New Astron.",
    volume = "6",
    pages = "79",
    year = "2001"
}

@article{Springel:2005mi,
    author = "Springel, Volker",
    title = "{The Cosmological simulation code GADGET-2}",
    eprint = "astro-ph/0505010",
    archivePrefix = "arXiv",
    doi = "10.1111/j.1365-2966.2005.09655.x",
    journal = "Mon. Not. Roy. Astron. Soc.",
    volume = "364",
    pages = "1105--1134",
    year = "2005"
}

@article{Yang:2022hkm,
    author = "Yang, Daneng and Yu, Hai-Bo",
    title = "{Gravothermal evolution of dark matter halos with differential elastic scattering}",
    eprint = "2205.03392",
    archivePrefix = "arXiv",
    primaryClass = "astro-ph.CO",
    doi = "10.1088/1475-7516/2022/09/077",
    journal = "JCAP",
    volume = "09",
    number = "09",
    pages = "077",
    year = "2022"
}

@article{Yang:2020iya,
    author = "Yang, Daneng and Yu, Hai-Bo and An, Haipeng",
    title = "{Self-Interacting Dark Matter and the Origin of Ultradiffuse Galaxies NGC1052-DF2 and -DF4}",
    eprint = "2002.02102",
    archivePrefix = "arXiv",
    primaryClass = "astro-ph.GA",
    doi = "10.1103/PhysRevLett.125.111105",
    journal = "Phys. Rev. Lett.",
    volume = "125",
    number = "11",
    pages = "111105",
    year = "2020"
}

@article{Garrison-Kimmel:2013yys,
    author = "Garrison-Kimmel, Shea and Rocha, Miguel and Boylan-Kolchin, Michael and Bullock, James and Lally, Jaspreet",
    title = "{Can Feedback Solve the Too Big to Fail Problem?}",
    eprint = "1301.3137",
    archivePrefix = "arXiv",
    primaryClass = "astro-ph.CO",
    doi = "10.1093/mnras/stt984",
    journal = "Mon. Not. Roy. Astron. Soc.",
    volume = "433",
    pages = "3539",
    year = "2013"
}

@article{vandenBosch:2018tyt,
    author = "van den Bosch, Frank C. and Ogiya, Go",
    title = "{Dark Matter Substructure in Numerical Simulations: A Tale of Discreteness Noise, Runaway Instabilities, and Artificial Disruption}",
    eprint = "1801.05427",
    archivePrefix = "arXiv",
    primaryClass = "astro-ph.GA",
    doi = "10.1093/mnras/sty084",
    journal = "Mon. Not. Roy. Astron. Soc.",
    volume = "475",
    number = "3",
    pages = "4066--4087",
    year = "2018"
}

@article{Burger:2021sep,
    author = "Burger, Jan D. and Zavala, Jes{\'u}s",
    title = "{Supernova-driven Mechanism of Cusp-core Transformation: an Appraisal}",
    eprint = "2103.01231",
    archivePrefix = "arXiv",
    primaryClass = "astro-ph.GA",
    doi = "10.3847/1538-4357/ac1a0f",
    journal = "Astrophys. J.",
    volume = "921",
    number = "2",
    pages = "126",
    year = "2021"
}

@ARTICLE{1911MNRAS..71..460P,
       author = {{Plummer}, H.~C.},
        title = "{On the problem of distribution in globular star clusters}",
      journal = {\mnras},
         year = 1911,
        month = mar,
       volume = {71},
        pages = {460-470},
          doi = {10.1093/mnras/71.5.460},
       adsurl = {https://ui.adsabs.harvard.edu/abs/1911MNRAS..71..460P}
}

\clearpage
\appendix

\section{End Matter}
\label{sec:em}

To quantify the competition between gravothermal collapse driven by self-interactions and energy injection from feedback, we estimate the SIDM thermalization timescale as~\cite{Tulin:2017ara}
\begin{equation}
    \label{eqn: scatter}
    t_{\rm th}\approx0.2\,{\rm Gyr}\left[\frac{0.1\mathrm{~M}_\odot/\mathrm{pc}^3}{\overline{\rho}}\right]\left[\frac{50\mathrm{~km/s}}{\left<v_\mathrm{rel}\right>}\right]\left[\frac{50\mathrm{~cm}^2/\mathrm{g}}{\sigma/m}\right],
\end{equation}
where $\overline{\rho}$ is the average density within a given radius, $\left<v_{\rm rel}\right>$ is the thermally averaged relative velocity of dark matter particles, and $\sigma/m$ is the self-interaction cross section per unit mass.

\begin{figure}[h]
\includegraphics[width=\linewidth]{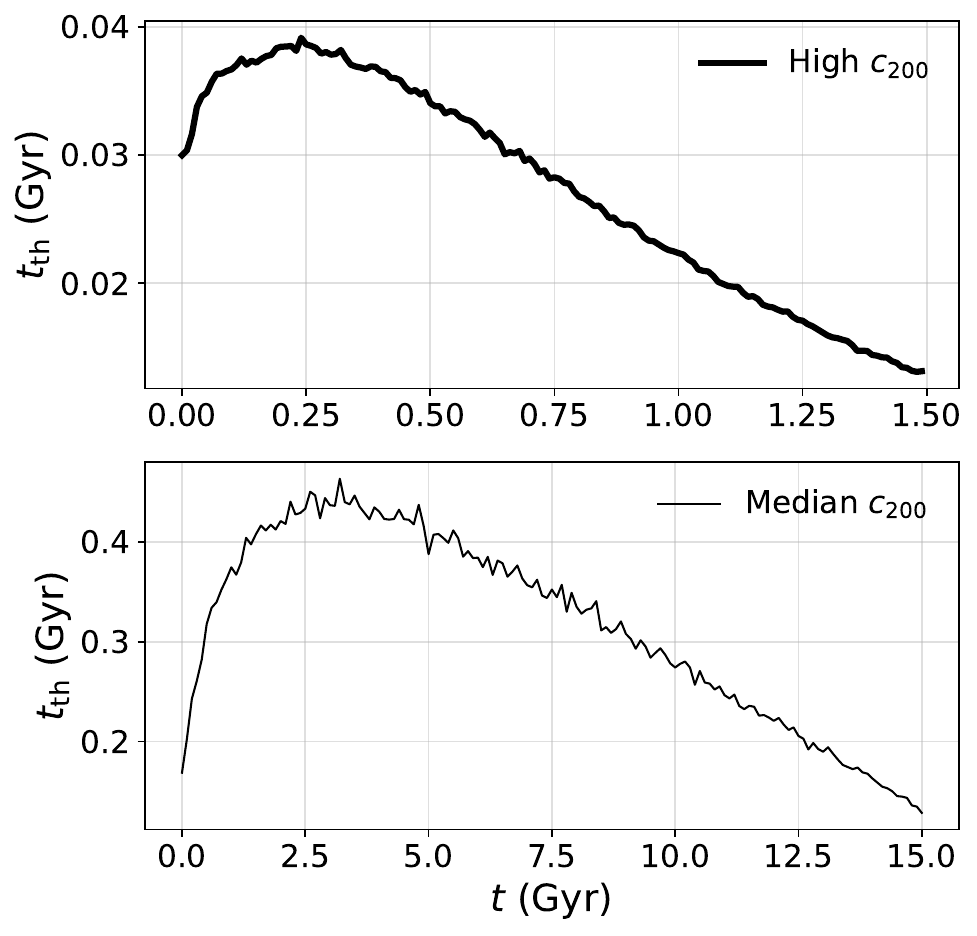}
\caption{Thermalization timescale during gravothermal evolution for the high-concentration (top) and median-concentration (bottom) halos in the SIDM-only simulations.}
\label{fig:thermal}
\end{figure}

Figure~\ref{fig:thermal} shows the evolution of $t_{\rm th}$ for the high-concentration (top) and median-concentration (bottom) halos in the SIDM-only runs. Here, $\overline{\rho}$ is computed within the inner $0.3\,{\rm kpc}$, $\left<v_{\rm rel}\right>=4\sigma_0/\sqrt{\pi}$ with $\sigma_0$ being the one-dimensional velocity dispersion in the same region, and $\sigma/m=50\,{\rm cm^2/g}$. In both halos, $t_{\rm th}$ initially increases during the core-expansion phase as the central density decreases, and then decreases once the halo enters core collapse and the central density rises.

For the high-concentration halo, $t_{\rm th}$ remains much shorter than the oscillation period of the baryonic potential $P=0.2\,{\rm Gyr}$, throughout the evolution, explaining why feedback only mildly delays collapse. In contrast, the median-concentration halo has $t_{\rm th}> P$ for most of its evolution, making it much more susceptible to feedback-driven heating and leading to a substantial delay of collapse. Fig.~\ref{fig:thermal} also shows that the gravothermal evolution of the high-concentration halo proceeds roughly an order of magnitude faster than that of the median-concentration halo.

\begin{figure}[h]
\includegraphics[width=\linewidth]{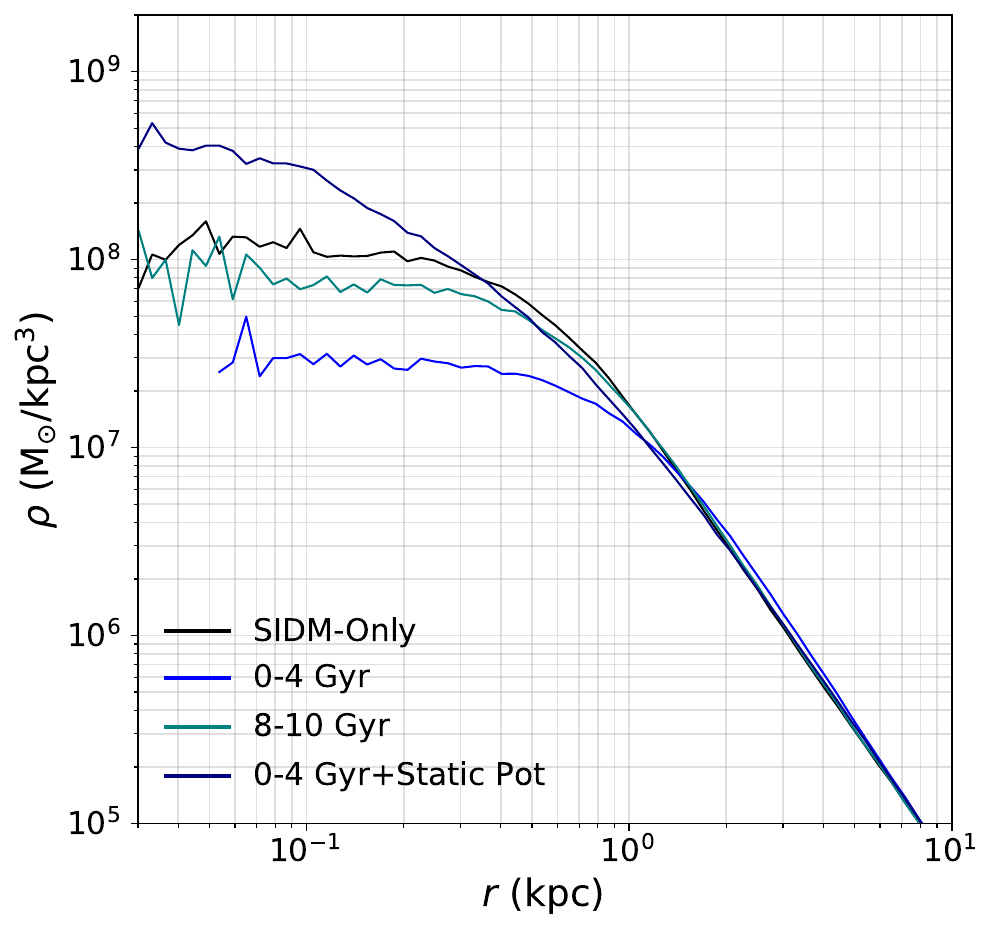}
\caption{The corresponding density profiles at $t=14\,\rm{Gyr}$ for the simulated cases in Fig.~\ref{fig:period}.}
\label{fig:denst14}
\end{figure}

In Fig.~\ref{fig:denst14}, we show the corresponding density profiles at $t=14\,\rm{Gyr}$ for the cases in Fig.~\ref{fig:period}. Different baryonic histories lead to distinct SIDM evolutionary trajectories, resulting in substantial variations in the final density profiles. Even for the same initial halo, the central densities can differ by an order of magnitude, depending on the feedback history during the SIDM evolution.

\end{document}